# Multiorbital tunneling ionization of the CO molecule

J. Wu[1,2], L. Ph. H. Schmidt[1], M. Kunitski[1], M. Meckel[1], S. Voss[1], H. Sann[1], H. Kim[1], T. Jahnke[1], A. Czasch[1], and R. Dörner[1,†]

[1]*Institut für Kernphysik, Goethe Universität, Max-von-Laue-Strasse 1, D-60438 Frankfurt, Germany*

[2]*State Key Laboratory of Precision Spectroscopy, East China Normal University, Shanghai 200062, China*

**Abstract**

We coincidently measure the molecular frame photoelectron angular distribution and the ion sum-momentum distribution of single and double ionization of CO molecules by using circularly and elliptically polarized femtosecond laser pulses, respectively. The orientation dependent ionization rates for various kinetic energy releases allow us to individually identify the ionizations of multiple orbitals, ranging from the highest occupied to the next two lower-lying molecular orbitals for various channels observed in our experiments. Not only the emission of a single electron, but also the sequential tunneling dynamics of two electrons from multiple orbitals are traced step by step. Our results confirm that the shape of the ionizing orbitals determine the strong laser field tunneling ionization in the CO molecule, whereas the linear Stark effect plays a minor role.

[†] Electronic address: doerner@atom.uni-frankfurt.de

Tunneling is one of the most prominent and important processes in the strong laser field ionization of atoms and molecules. It underlies a wide range of fundamental physical phenomena such as the emission of coherent radiation in the x-ray region which is the basis for attosecond science [1,2], the laser induced electron diffraction and holography [3] and nosequential double ionization [4]. In general, the molecular ionization dynamics in a strong field can successfully be described by the molecular Ammosov-Delone-Krainov (MO-ADK) theory [5,6] and the strong field approximation (SFA) calculation [7]. More recently also direct solutions of the time dependent Schrödinger equation became available for molecules [8,9]. Initially most studies concluded that the angular dependent ionization rate maps the electron density profile or the shape of the ionizing orbital [5,7,10]. More recently however this has been questioned [11-19]. Deviations of the observed rates from the expectation based on the shape of the highest occupied molecular orbital (HOMO) have been attributed to an interplay of the coordinate-space and momentum-space features of the ionizing orbital [12], the combined contributions of multiple orbitals [13-16], and to the modification of the ionization potential by the linear Stark effect due to the orbital dipole [17-19]. The linear Stark effect raises or reduces the ionization potential when the laser field vector is parallel or anti-parallel, respectively, to the orbital dipole. Thus, for instance, in the case of the OCS molecule [17], the linear Stark effect reverses the orientation dependent ionization rate as compared to the anticipation of the traditional molecular ADK theory.

For carbon monoxide (CO), the influence of the linear Stark effect on the orientation dependent ionization rate is still a hot topic of controversial debate [20-25]. The Stark-corrected MO-ADK [17,21,24] indicates a maximum ionization rate when the laser field points from nuclei O to C which is opposite to the traditional MO-ADK theory; while the Stark-corrected molecular SFA [18,24] predicts a maximum ionization rate when the laser field points from nuclei C to O. The recent experimental measurements [23,24] of the emission direction of the dissociative ions from multiply ionized CO by using intense linearly polarized two-color pulses demonstrate that the linear Stark effect plays a minor role and the ionization rate is dominated by the orbital profile. However no coincident measurement of the angular distribution of the emitted electron and ion has been performed so far, which would serve as the most direct approach to reveal the orientation dependent ionization rate [17]. Even more challenging than predicting and measuring single electron emission is tracing the multielectron emission dynamics involving complex transitions in a many-particle system and it is completely open how far the simplified concept of the single active electron model can be taken.

In this paper, we measure the orientation dependent single and double ionization rate of the CO molecule by using single-color circularly and elliptically polarized femtosecond laser pulses. Not only the emission of single electron, but also the sequential tunneling dynamics of two electrons from multiple orbitals are individually identified and angularly resolved by observing the kinetic energy release (KER) dependent molecular-frame photoelectron angular distribution (MFPAD) and the asymmetry of the ion sum-momentum distribution of the exploded singly and doubly ionized CO molecules. Our results suggest that for all of the identified multiple orbitals and for the first and second ionization step the linear Stark effect due to an orbital dipole plays a minor role in the strong field tunneling ionization of CO molecule. They are instead governed mainly by the shape of the respective orbital.

The measurements were performed using a reaction microscope of COLd Target Recoil Ion Momentum Spectroscopy (COLTRIMS) [26,27], where the molecules from a collimated supersonic gas jet were ionized by a 35-fs laser pulse with a central wavelength of 790 nm and a repetition rate of 8 kHz. The 3D momenta of the coincidently emitted electrons and ions from the ionization were reconstructed from the detected times-of-flight and positions of the charged particles. As shown in Fig. 1, the polarization of the input laser pulse was adjusted to be circular or elliptical before it entered the interaction chamber. The major polarization axis of the ellipse was set to be along the y-axis. The electron freed by the laser field pointed along the y-axis acquired a final momentum along the z-axis (i.e. the time-of-flight direction of the spectrometer which provides a best momentum resolution) due to the rotation of the applied laser field. This "angular streaking" concept [1,28,29] when close-to-circular or elliptically polarized pulses are used for ionization allows to suppress the recollision process and map the instantaneous magnitude and direction of the laser field vector at the moment of ionization to the magnitude and direction of the final momentum of the emitted electron. The molecular axis and instantaneous ionization laser field vector were deduced from the measured momenta of the coincident ions and electrons. The geometric rotation of the molecule during the ionization is negligible for the 35-fs laser pulse used in the experiment. This enables the study of the orientation dependent ionization rate without pre-orientation of the target molecule, as firstly demonstrated by Staudte *et al.* for the angular-dependent ionization of $H_2$ [30].

Figure 2(a) shows the KER dependent MFPAD of the single ionization induced dissociative channel of $C^++O$ by using circularly polarized single-color femtosecond laser pulses [Fig. 1(a)]. In the analysis, the orientation of the molecule is fixed along the y-axis by rotating the emission direction of $C^+$ to the positive direction of the y-axis (i.e. $\phi_e=0°$). The angular distribution of the coincidently recorded electron is then reconstructed in the molecular frame, i.e. MFPAD. As shown in Fig. 2(b), the KER distribution of the $C^++O$ channel shows two peaks separated by a minimum around 0.75 eV. The MFPADs integrated over different KER ranges are plotted in Fig. 2(c). The high KER peak (>0.75 eV) shows a significant asymmetry in the angular distribution as compared to the low KER peak (<0.75 eV). The count rate around $\phi_e=90°$ is 16.8% higher than around $\phi_e=-90°$ indicating a higher ionization rate of CO molecule when the laser field points from C to O. It is in agreement with the conclusions drawn from recent measurements of the emission directions of the ions [23,24] in phase controlled two-color linearly polarized pulses. This indicates that the ionization of CO is predominated by the orbital profile rather than the linear Stark effect, probably because of the small orbital dipole moment, e.g. ~1.7 a.u. for the HOMO [21]. The predicted angular-dependent ionization rates of CO simulated by using the MO-ADK and SFA as well as their corrections with the Stark effect are shown in Fig. 2(d) (adapted from Ref. [24]), where our measured MFPAD (rotated by -90° to take into account the angular-streaking effect of the circularly polarized pulse) for the high KER peak of the $C^++O$ is also presented for comparison. The SFA calculation with Stark effect correction [18,24] seems to be a good approximation for the strong field tunneling ionization of CO under the current experimental conditions.

Beyond the HOMO, the next lower-lying orbitals (HOMO-1 or HOMO-2) can also contribute to the single ionization of the molecules [2,13-16,31]. For the CO molecule exposed to a strong laser field, the ionic bound states $X^2\Sigma^+$ or $A^2\Pi$ of $CO^+$ are firstly populated by removing an electron from HOMO or HOMO-1 of the neutral CO [see Fig. 1(d)]. It is then transferred to the repulsive potential energy curve by the laser-induced coupling and leads to the observed dissociative channel of $C^++O$ [15,31]. The recollision excitation plays a minor role in our circularly polarized laser pulse, and only a dissociative channel of $C^++O$ with KER of less than 2 eV is observed [Fig. 2(b)] which is mainly produced by the laser-induced coupling [15]. The KER dependent MFPAD distribution presented in Fig. 2 allows us to identify the ionizations from multiple orbitals for this single electron ionization channel. As illustrated in Fig. 1(c), the molecular orbital of HOMO-1 is more homogeneously distributed between the nuclei as compared with HOMO. Since the ionization is predominated by the geometric profile of the ionizing orbital, a larger asymmetry in the orientation dependent ionization rate is expected for HOMO than HOMO-1. As illustrated in Fig. 2(c), the MFPAD of the high KER peak shows a much more significant asymmetry compared to the low KER peak. It indicates that the high and low KER peaks are produced by the ionization of the HOMO and HOMO-1, respectively. The HOMO-1 has $\Pi$ symmetry and a nodal plane along the molecular axis. Other than in previous experiments on $O_2$ [11] and HCl [14] we do not see this node in the MFPADs. The $X^2\Sigma^+$ ionic state has a smaller equilibrium internuclear distance than the $A^2\prod$ one [see Fig. 1(d)], which results in a higher KER when it is further coupled to the repulsive potential curve. Additionally, the emitted electron correlated with the low KER peak is measured to have a higher kinetic energy (~14.9 eV) than that correlated with the high KER peak (~13.9 eV). This is consistent with the fact that the HOMO-1 has a higher bound energy (16.6 eV) than the HOMO (14.1 eV), thus a higher laser field intensity is required to remove the electron from the HOMO-1, leading to a larger final momentum or kinetic energy of the freed electron [28,29]. Even the energy level of HOMO-1 is ~2.5 eV lower than that of HOMO, the contribution of HOMO-1 to the dissociative single ionization process is significant (~30% of the total signal measured in our experiment).

The ionization of the second electron from multiple orbitals is also observed for the double ionization induced Coulomb exploded channel of $C^++O^+$. For ejection of two electrons their sum-momentum has its mirror image in the sum-momentum of the two ions, which is readily measured with high resolution in our setup. As illustrated in Fig. 1(b), an elliptically polarized single-color pulse with major polarization axis along the y-axis is used to doubly ionize the CO molecule. Similar to the case of linear polarization [10,31], we find that those molecules are predominantly doubly ionized which are aligned parallel to the major polarization axis of our elliptically polarized laser pulse. For the following analysis we have selected only ions emitted within a 45º cone with respect to the major polarization axis (y-axis).

We identify the multiple-orbitals double ionization of the CO molecule by analyzing the asymmetric ionization rate when the laser field points from nuclei C to O or O to C. This asymmetric ionization along the molecular axis is the crucial question we are trying to address here, and which is also currently one of the controversial debates in the literature. Please note that the single-color elliptically polarized pulse does not introduce any bias to this asymmetric ionization dynamics. More importantly, in the elliptically polarized pulse, the electron is mainly freed when the laser field points along the major axis which ends with a final momentum along the minor axis due to the angular streaking of the rotating laser field. This serves as the heart to distinguish the direction of the laser field vector at the moment of ionization and to reveal the asymmetric ionization dynamics by deconvoluting the ion sum-momentum along the minor polarization axis.

Figure 3(b) displays the KER dependent ion sum-momentum distribution of the exploded double ionization channel of $C^++O^+$ along the z-axis ($pz_{sum}$). The three-peak structure of $pz_{sum}$ reflects the sum-momentum of the sequentially freed two electrons according to the momentum conservation [28,32]. The measured ion sum-momentum distribution thus can be used to study the ionization dynamics of the coincidently emitted electrons. For the anti-clockwise rotating elliptical field [Fig. 1(b)], the electron freed by the laser field points to the negative (-y) or positive (+y) directions of the y-axis and receives a final momentum of positive ($pz_e>0$) or negative ($pz_e<0$) along z-axis, respectively. For our elliptically polarized pulse with an ellipticity of $\varepsilon$~0.7, the double ionization predominantly proceeds through a sequential process [28,29] when the laser field points along the major polarization axis, thus no correlation between the two freed electrons is expected. The center peak in the observed three peak structure of the sum-momentum (Fig. 3) results from double ionization events where both electrons are ejected to opposite direction canceling their momenta. The two side peaks are filled by events where both electron escape to the same side of the molecule. We can model this in more detail by assuming a Gaussian distribution of the electron momentum centered at $pz_{ek}$, i.e. $A_k/[\sigma_k\sqrt{(2\pi)}]\times \exp[-0.5(pz-pz_{ek})^2/\sigma^2_k]$ (k=1 or 2 accounts for the first or second electron), the recoil ion sum-momentum (i.e. the reverse of sum-momentum of the sequentially freed two electrons) can be expressed as

$$pz_{sum12}(pz) = \frac{1}{\sqrt{2\pi}} \sum_{i,j=+,-} \frac{A_{12ij}}{\sigma_{12ij}} \exp[-0.5(pz-pz_{sumij})^2/\sigma^2_{12ij}] , \qquad (1)$$

where $A_{12ij}=A_{1i}\times A_{2j}$, $\sigma^2_{12ij}=\sigma^2_{1i} + \sigma^2_{2j}$, and $pz_{sumij}= -(ipz_{e1} +jpz_{e2})$ accounts for four possible ion sum-momenta. For our laser parameters, the magnitudes of $pz_{e1}$ and $pz_{e2}$ are very similar and hence ($pz_{e1}$ - $pz_{e2}$) ~ (-$pz_{e1}$ + $pz_{e2}$) ~ 0, leading to a single peak of $pz_{sum}$ around zero for these two cases. The sides peaks of positive [$pz_{sum}$=-(-$pz_{e1}$ - $pz_{e2}$) >0] or negative [$pz_{sum}$=-($pz_{e1}$ + $pz_{e2}$) <0] account for the cases that both two electrons are freed when the laser field points to +y or -y, respectively.

For our multi-cycle single-color laser pulse, the overall ionization probabilities when the laser field points to +y and –y are the same, and therefore a symmetric distribution of $pz_{sum}$ is resulted as displayed in Fig. 3(b). However, as plotted in Figs. 3(c) and 3(d), highly asymmetric distributions of $pz_{sum}$ are observed when we select the $C^+$ ion emits to +y ($py_{C+}$>0) or –y ($py_{C+}$<0), according to the initial orientation of CO with the C atom pointing to +y or –y, respectively. The higher side peak at $pz_{sum}$>0 (or $pz_{sum}$<0) for $py_{C+}$<0 (or $py_{C+}$>0) suggests a higher ionization rate when the laser field is pointing to +y (or -y), which is opposite to the orientation direction of atom C. In agreement with our conclusions for dissociative single ionization discussed above and the pioneering measurements of the ion emission direction by using phase controlled two-color laser pulses [23,24], the double ionization of CO is more likely to occur when the laser field is pointing from C to O. We have cross checked our experiment by inverting the polarization rotating direction. This leads as expected to an inversion of the asymmetry (data not shown here).

In order to get quantitative insights into the orientation dependent double ionization of the CO molecule, we now fit the asymmetric ion sum-momentum distribution with the function of $pz_{sum12}$ [equation (1)]. The solid curves in Fig. 3(e) show the fits with the following parameters: $pz_{e1}$=0.91 a.u., $pz_{e2}$=1.38 a.u., $A_{1\pm}/A_{1\mp}$=1.41, and $A_{2\pm}/A_{2\mp}$=1.65. The significant difference between the momenta of electron 1 and 2 indicates that a higher laser field is required to remove the second electron which is more tightly bound [28]. The second ionization step with a larger asymmetric parameter is more sensitive to the orientation of molecular CO than the first ionization step.

Similar to single ionization, the Coulomb exploded double ionization channel

C$^+$+O$^+$ features two distinct KER peaks separated with a minimum around 7.5 eV [Fig. 3(a)] and shows different ion sum-momentum distributions [Fig. 3(b)]. This KER dependence can be noticed more clearly by selecting the ions when the C$^+$ emits to +y [py$_{C+}$>0, Fig. 3(c)] or −y [py$_{C+}$<0, Fig. 3(d)] and integrating over different KER ranges [KER<7.5 eV in Fig. 3(f), and KER>7.5 eV in Fig. 3(g)]. Different asymmetries in the ion sum-momentum distributions are observed for the low and high KER peaks. For the low KER peak, the fits using equation (1) [solid curves in Fig. 3(f)] return: pz$_{e1}$=0.9 a.u., pz$_{e2}$=1.33 a.u., A$_{1\pm}$/A$_{1\mp}$=1.27, and A$_{2\pm}$/A$_{2\mp}$=1.49. For the high KER peak, the fits with the function of pz$_{sum12}$ [solid curves in Fig. 3(g)] return: pz$_{e1}$=0.92 a.u., pz$_{e2}$=1.43 a.u., A$_{1\pm}$/A$_{1\mp}$=1.54, and A$_{2\pm}$/A$_{2\mp}$=1.81. For both the first and second ionization steps, the high KER peak shows a larger asymmetry in pz$_{sum}$ than the low KER peak. Meanwhile, the momentum of the second electron correlated with the high KER peak is larger than that of the low KER peak.

In addition to the HOMO and HOMO-1, the HOMO-2 may also contribute to double ionization. As illustrated in Fig. 1(c), the electron wavefunction is mainly located on the C side for HOMO-2. The biggest asymmetry in the ionization rate is therefore expected when an electron is removed from HOMO-2. The pathways for producing C$^+$+O$^+$ are illustrated in Fig. 1(d). Starting from the removal of the first electron from HOMO (populates the bound state X$^2\Sigma^+$ of CO$^+$) or HOMO-1 (populates the bound state A$^2\Pi$ of CO$^+$), the removal of the second electron from HOMO-2 or HOMO leads to the population of the bound states $^3\Sigma^+$ or X$^3\Pi$ of CO$^{2+}$, respectively, which then dissociates through the crossing repulsive potential curve (e.g. $^3\Sigma^-$). Due to its short equilibrium internuclear distance [see Fig. 1(d)], the dissociation initialized from the state of $^3\Sigma^+$ will lead to a higher KER as compared with that from X$^3\Pi$. It suggests that the observed high and low KER peaks in the C$^+$+O$^+$ channel are launched from the $^3\Sigma^+$ and X$^3\Pi$ ionic states of CO$^{2+}$, respectively. The Franck Condon overlap will strongly favor vertical transitions between curves of similar equilibrium distances between X$^2\Sigma^+$ and $^3\Sigma^+$ or between A$^2\Pi$ and X$^3\Pi$. This is consistent with the above mentioned observation: the removal of the second electron from HOMO-2 (leading to the vertical transition from $X^2\Sigma^+$ to $^3\Sigma^+$) gives rise to a larger electron momentum $pz_{e2}$ (a higher laser field intensity is required to remove the tightly bound electron from HOMO-2) than the removal of the second electron from the HOMO (leading to the vertical transition from $A^2\Pi$ to $X^3\Pi$). We thus identify two ionization pathways to produce the exploded double ionization channel of $C^++O^+$: the sequential removals of the first and second electrons from the HOMO-1 and HOMO lead to the observed $C^++O^+$ signal in the low KER range; while the high KER range signal of $C^++O^+$ mainly proceeds through the removals of the first and second electrons from the HOMO and HOMO-2, respectively.

In summary, we identified the individual removal of the first and second electrons from multiple orbitals for the exploded single and double ionization of the CO molecule, featuring with the KER dependent MFPADs and asymmetric ion sum-momentum distributions. The influence of the linear Stark effect of the orbital dipole is found to be weak for the orientation dependent ionization of molecular CO for all orbitals and both ejected electrons. The tunneling ionization is dominated by the geometric profile of the ionizing orbitals. In agreement with the pioneering experiments [2,13-16,31], our results suggest that tunneling ionization of multiple orbitals is fairly general and essential to understand the strong laser field dynamics of small molecules.

**Acknowledgement**: J.W. acknowledges support by the Alexander von Humboldt Foundation, support by a Koselleck Project of the Deutsche Forschungsgemeinschaft is acknowledged.

**References:**

[1] P. Eckle *et al.*, Science **322**, 1525 (2008).

[2] O. Smirnova *et al.*, Nature **460**, 972 (2009).

[3] M. Meckel *et al.*, Science **320**, 1478 (2008).

[4] R. Dörner et al, Adv. At. Mol. Phys. **48**, 1 (2002).

[5] X. M. Tong, Z. X. Zhao, and C. D. Lin, Phys. Rev. A **66**, 033402 (2002).

[6] O. I. Tolstikhin, T. Morishita, and L. B. Madsen, Phys. Rev. A **84**, 053423 (2011).

[7] J. Muth-Böhm, A. Becker, and F. H. M. Faisal, Phys. Rev. Lett. **85**, 2280 (2000).

[8] N. Takemoto and A. Becker, Phys. Rev. Lett. **105**, 203004 (2010).

[9] M. Abu-samha and L. B. Madsen, Phys. Rev. A **81**, 033416 (2010).

[10] A. S. Alnaser *et al.*, Phys. Rev. A **71**, 031403(R) (2005).

[11] D. Pavičić *et al.*, Phys. Rev. Lett. **98**, 243001 (2007).

[12] R. Murray, M. Spanner, S. Patchkovskii, and M. Yu. Ivanov, Phys. Rev. Lett. **106**, 173001 (2011).

[13] B. K. McFarland, J. P. Farrell, P. H. Bucksbaum, and M. Gühr, Science **322**, 1232 (2008).

[14] H. Akagi *et al.*, Science **325**, 1364 (2009).

[15] I. Znakovskaya *et al.*, Phys. Rev. Lett. **103**, 103002 (2009)

[16] C. Wu *et al.*, Phys. Rev. A. **83**, 033410 (2011).

[17] L. Holmegaard *et al.*, Nature Phys. **6**, 428 (2010).

[18] D. Dimitrovski, C. P. J. Martiny, and L. B. Madsen, Phys. Rev. A **82**, 053404 (2010).

[19] D. Dimitrovski *et al.*, Phys. Rev. A **83**, 023405 (2011).

[20] S. De *et al.*, Phys. Rev. Lett. **103**, 153002 (2009).

[21] M. Abu-samha and L. B. Madsen, Phys. Rev. A **82**, 043413 (2010).

[22] K. J. Betsch, D. W. Pinkham, and R. R. Jones, Phys. Rev. Lett. **105**, 223002 (2010).

[23] H. Ohmura, N. Saito, and T. Morishita, Phys. Rev. A **83**, 063407 (2011).

[24] H. Li *et al.*, Phys. Rev. A **84**, 043429 (2011).

[25] X. Zhu *et al.*, Opt. Express 19, 24198 (2011).

[26] R. Dörner *et al.*, Phys. Rep. **330**, 95 (2000).

[27] J. Ullrich *et al.*, Rep. Prog. Phys. **66**, 1463 (2003).

[28] A. N. Pfeiffer *et al.*, Nat. Phys. **7**, 428 (2011).

[29] J. Wu *et al.*, Phys. Rev. Lett. **108**, 043002 (2012).

[30] A. Staudte *et al.*, Phys. Rev. Lett. **102**, 033004 (2009).

[31] S. De *et al.*, Phys. Rev. A **84**, 043410 (2011).

[32] C. M. Maharjan *et al.*, Phys. Rev. A **72**, 041403(R) (2005).

**Figure Captions**

**Figure 1** (Color online) The sketch of the anti-clockwise rotating (a) circularly and (b) elliptically polarized laser field and the angular streaking concept in our coordinate system. (c) The highest occupied three orbitals involved in the single and double ionizations of CO molecule in our experiments, where the nuclei O and C lie on the left and right sides, respectively. (d) The potential energy surfaces of $CO^+$ and $CO^{2+}$ interested in this work (adapted from Ref. [31]). The $X^2\Sigma^+$ or $A^2\Pi$ state of $CO^+$ is populated by removing an electron from HOMO or HOMO-1 of the neutral CO, which transits to the $^3\Sigma^+$ or $X^3\Pi$ state of $CO^{2+}$ by removing the second electron from HOMO-2 or HOMO, respectively.

**Figure 2** (Color online) (a) The KER dependent molecular frame angular distribution of the photoelectron emitted from $C^++O$, and (b) its integrations over different KER ranges. (c) The KER distribution of the single ionization induced dissociation channel of $C^++O$. The CO molecule is orientated with C points to the positive direction of the y-axis (i.e. $\phi_e=0°$). The $\phi_e=+90°$ corresponds to electron tunneling when the electric field points from C towards O, i.e. the Carbon is on the down hill side of the potential. The electrons are then angularly streaked towards $\phi_e=+90°$. The intensity of the anti-clockwise rotating circularly polarized pulse is estimated to be $\sim 4.0\times 10^{14}$ W/cm$^2$. (d) The predicted angular-dependent ionization rates of CO simulated by using the MO-ADK and SFA as well as their corrections with the Stark effect (adapted from Ref. [24]), which are normalized and compared with our experimental data (the measured angular distribution is rotated by -90° to take into account the angular-streaking effect).

**Figure 3** (Color online) (a) The KER distribution of the exploded double ionization channel of $C^++O^+$. (b-d) The KER dependent ion sum-momentum distributions of the $C^++O^+$ channel along the minor polarization axis (z-axis, $pz_{sum}$) of the elliptically polarized pulse. The condition of the $C^+$ emits to the positive ($py_{c+}>0$) or negative ($py_{c+}<0$) direction of the y-axis is applied in (c) and (d), respectively. The corresponding ion sum-momentum distributions integrated over different KER ranges

are plotted in (e-g). The solid curves show the fits of the measured three-peak structures by considering the convolution of two sequentially emitted electrons according to the function of $pz_{sum12}$ (see text). Only the molecules orientated within a cone of 45˚ around the major polarization axis (y-axis) are selected. The intensity of the anti-clockwise rotating elliptically polarized pulse with an ellipticity of ε~0.7 is estimated to be ~$2\times10^{15}$ W/cm$^2$.

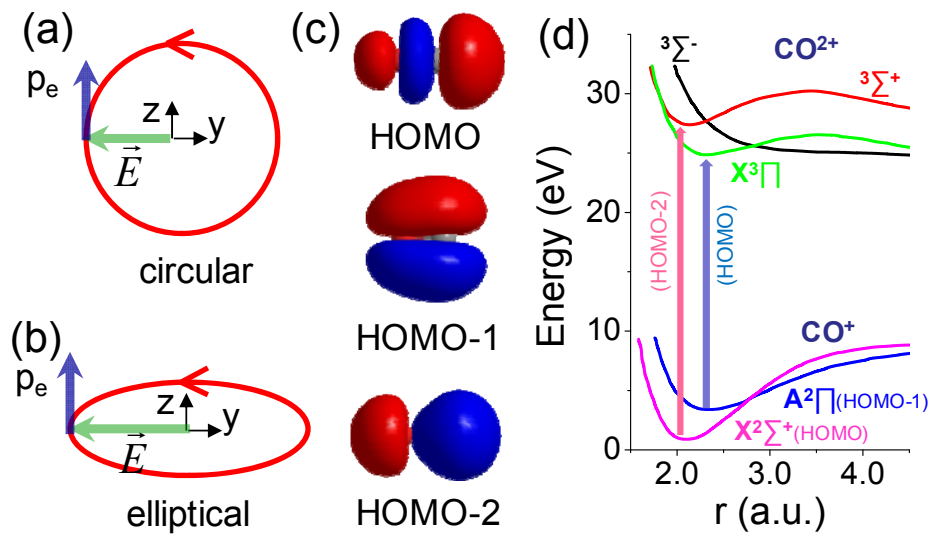

**Figure 1** (Color online) The sketch of the anti-clockwise rotating (a) circularly and (b) elliptically polarized laser field and the angular streaking concept in our coordinate system. (c) The highest occupied three orbitals involved in the single and double ionizations of CO molecule in our experiments, where the nuclei O and C lie on the left and right sides, respectively. (d) The potential energy surfaces of $CO^+$ and $CO^{2+}$ interested in this work (adapted from Ref. [31]). The $X^2\Sigma^+$ or $A^2\Pi$ state of $CO^+$ is populated by removing an electron from HOMO or HOMO-1 of the neutral CO, which transits to the $^3\Sigma^+$ or $X^3\Pi$ state of $CO^{2+}$ by removing the second electron from HOMO-2 or HOMO, respectively.

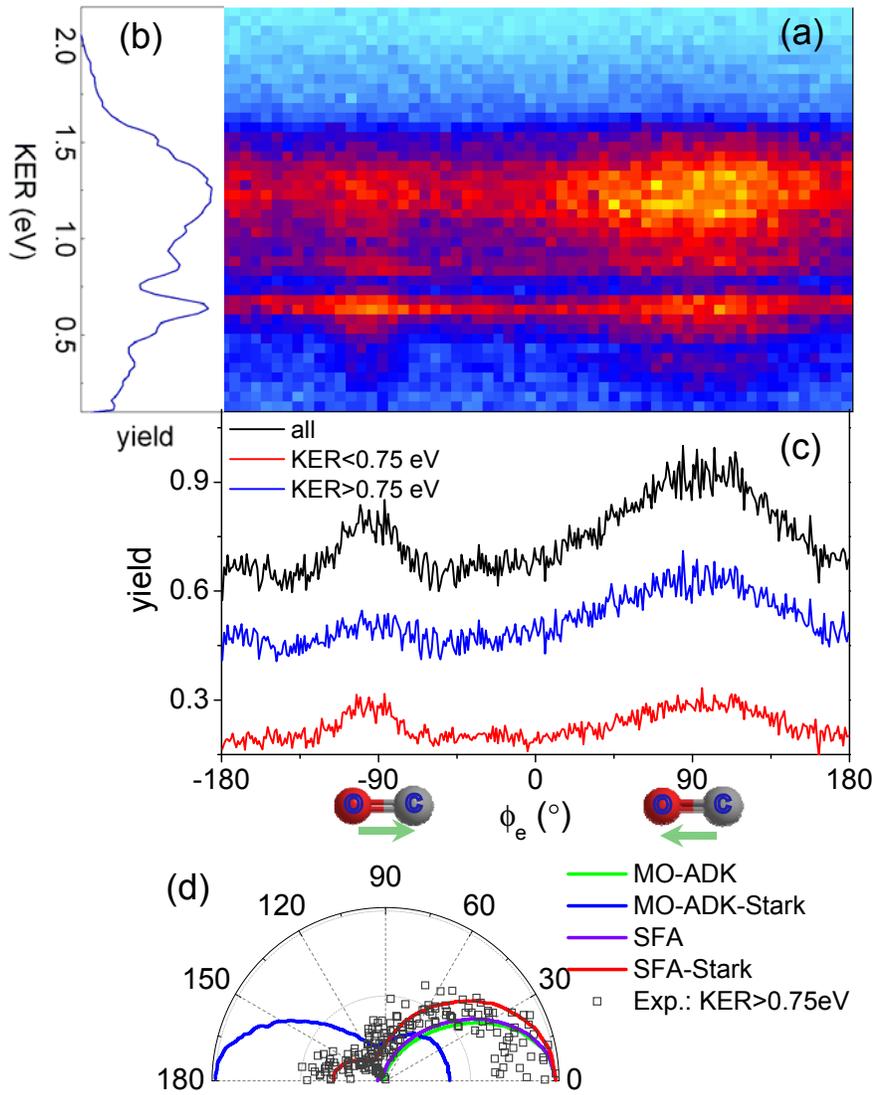

**Figure 2** (Color online) (a) The KER dependent molecular frame angular distribution of the photoelectron emitted from $C^++O$, and (b) its integrations over different KER ranges. (c) The KER distribution of the single ionization induced dissociation channel of $C^++O$. The CO molecule is orientated with C points to the positive direction of the y-axis (i.e. $\phi_e=0°$). The $\phi_e=+90°$ corresponds to electron tunneling when the electric field points from C towards O, i.e. the Carbon is on the down hill side of the potential. The electrons are then angularly streaked towards $\phi_e=+90°$. The intensity of the anti-clockwise rotating circularly polarized pulse is estimated to be $\sim 4.0\times 10^{14}$ W/cm$^2$. (d) The predicted angular-dependent ionization rates of CO simulated by using the MO-ADK and SFA as well as their corrections with the Stark effect (adapted from Ref. [24]), which are normalized and compared with our experimental data (the measured angular distribution is rotated by -90° to take into account the angular-streaking effect).

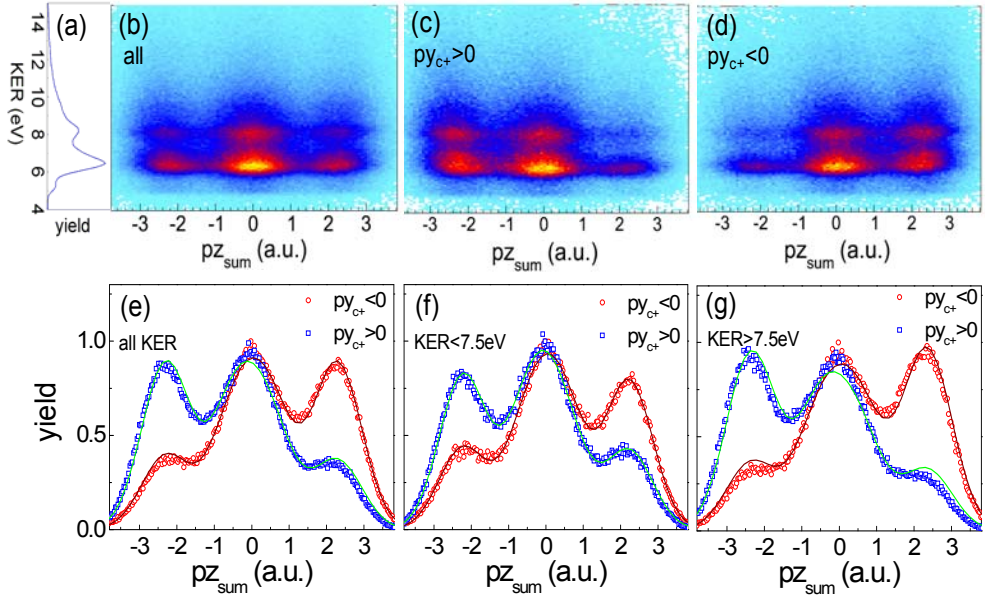

**Figure 3** (Color online) (a) The KER distribution of the exploded double ionization channel of $C^++O^+$. (b-d) The KER dependent ion sum-momentum distributions of the $C^++O^+$ channel along the minor polarization axis (z-axis, $pz_{sum}$) of the elliptically polarized pulse. The condition of the $C^+$ emits to the positive ($py_{c+}>0$) or negative ($py_{c+}<0$) direction of the y-axis is applied in (c) and (d), respectively. The corresponding ion sum-momentum distributions integrated over different KER ranges are plotted in (e-g). The solid curves show the fits of the measured three-peak structures by considering the convolution of two sequentially emitted electrons according to the function of $pz_{sum12}$ (see text). Only the molecules orientated within a cone of 45° around the major polarization axis (y-axis) are selected. The intensity of the anti-clockwise rotating elliptically polarized pulse with an ellipticity of ε~0.7 is estimated to be ~$2\times10^{15}$ W/cm$^2$.